\documentclass[aps,superscriptaddress,twocolumn,twoside,floatfix,pra,nofootinbib,a4paper]{revtex4-2}
\usepackage{times}
\usepackage{epsfig}
\usepackage{amsfonts}
\usepackage{amsmath}
\usepackage{amssymb}
\usepackage{amsthm}
\usepackage{color}
\usepackage{multirow}
\usepackage[normalem]{ulem}
\newcommand{\stkout}[1]{\ifmmode\text{\sout{\ensuremath{#1}}}\else\sout{#1}\fi}
\usepackage{latexsym}
\usepackage{mathrsfs}
\usepackage{natbib}
\usepackage{verbatim}
\usepackage[T1]{fontenc}
\usepackage{float}
\usepackage{subfig}
\usepackage{inputenc}

\usepackage{ragged2e}
\justifying

\usepackage{graphicx}
\usepackage{xcolor}

\usepackage[colorlinks=true,linkcolor=blue,citecolor=magenta,urlcolor=blue]{hyperref}

\begin{document}


\title{Experimental demonstration of full network nonlocality in the bilocal scenario}


\author{Emil H\aa kansson}
\affiliation{Department of Physics, Stockholm University, S-10691 Stockholm, Sweden}
\affiliation{Hitachi Energy Research, Forskargränd 7, 72219 V\"aster\aa s, Sweden}

\author{Amélie Piveteau}
\affiliation{Department of Physics, Stockholm University, S-10691 Stockholm, Sweden}

\author{Sadiq Muhammad}
\affiliation{Department of Physics, Stockholm University, S-10691 Stockholm, Sweden} 

\author{Mohamed Bourennane}
\affiliation{Department of Physics, Stockholm University, S-10691 Stockholm, Sweden}

\date{\today}

\maketitle

\textbf{Quantum correlation between nodes in a network which consist of several independent sources of entanglement and in multipartite  entanglement systems are important for general understanding of the nature of nonlocality, quantum information processing and communication. In previous years, demonstrations of network nonlocality for bilocal scenarios have been in the focus. Yet, it has been found that the seminal protocols do not certify entanglement between end nodes, which otherwise require all sources in the network to be of nonlocal nature. This has motivated for development of even stronger concept, called full network nonlocality. Here, we  experimentally demonstrate full network nonlocal correlations in a network. Specifically, we use two pairs of polarization entangled qubits created by two separate and independent entanglement sources and a partial Bell state measurement. Our results may pave the way for the  use  of  nonlocality test in quantum communication protocols with a full network nonlocality certification.}

\begin{figure}[t!]
  \centering
  \subfloat[][]{\includegraphics[width=0.63\columnwidth]{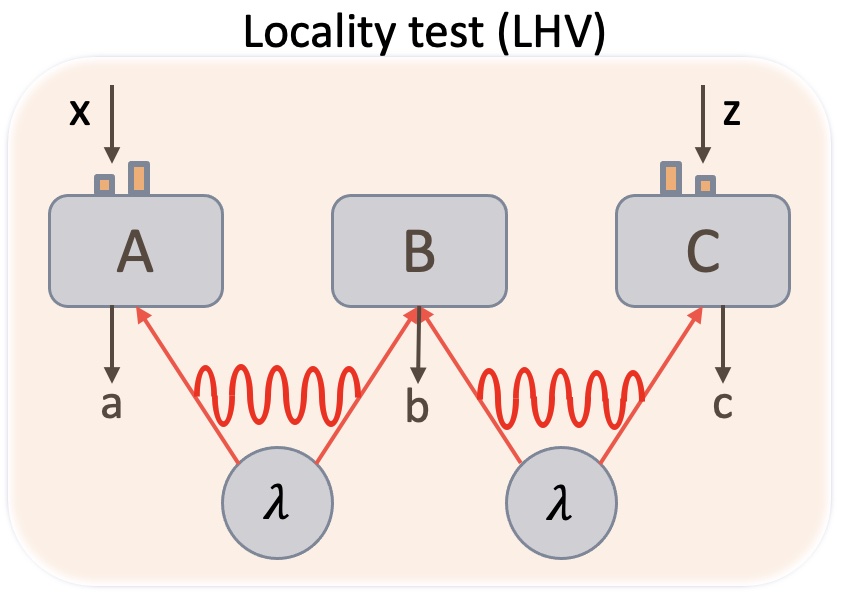}
  \label{fig:a}} \\ 
  \subfloat[][]{\includegraphics[width=0.63\columnwidth]{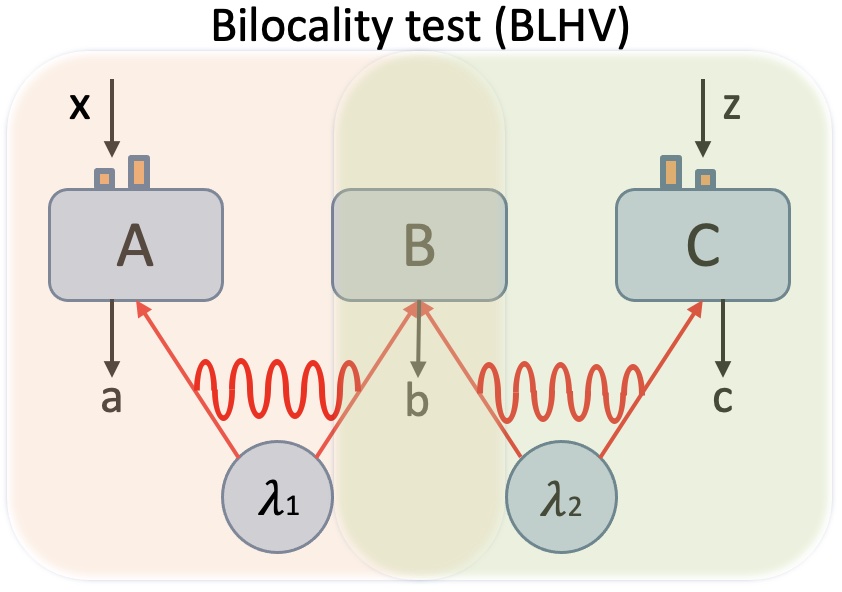}}\\
  \subfloat[][]{\includegraphics[width=0.63\columnwidth]{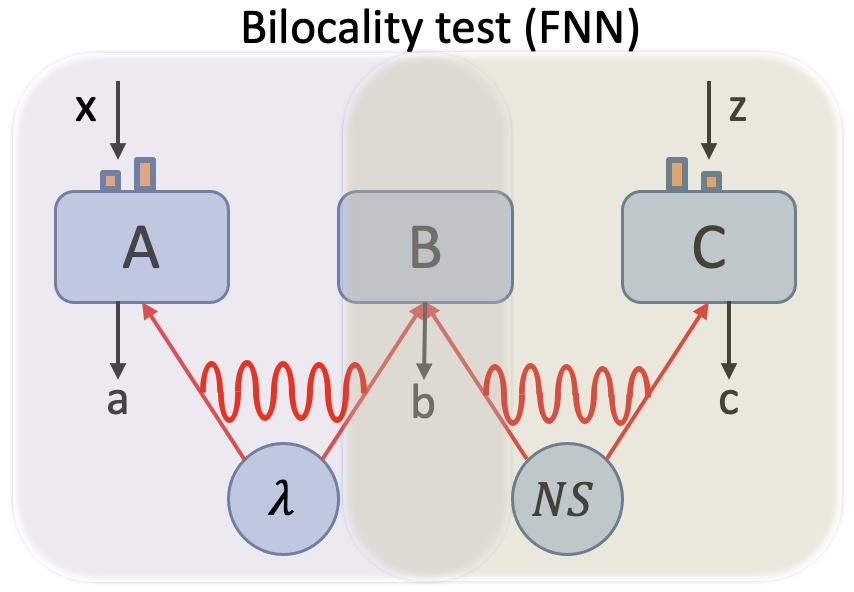}}
\caption{ \emph{Bilocal quantum network}. Nodes, blue rectangles where  A, B, and  C indicated Alice 's , Bob's, and  Charlie's  nodes respectively. $(x, a)$, $(y,b)$ , and  $(z,c)$ are  the  input and  output  of  Alice , Bob, and  charlie   respectively. The  parties are connected by red arrows which symbolize a sharing of a common source. (a) LHV model and its event-ready Bell test, where the region of influence is steered by one hidden variable $\lambda$. 
(b) BLHV model with two different local hidden variables ($\lambda_1$ and $\lambda_2$) represented as two different regions of influence (different shadings). 
(c)  FNN scenario. The combination with one local hidden variable and an non-local source ($NS$) (represented by two different shadings) given rise to local full network model.}\label{Fig1}
\end{figure} 

Quantum entanglement is one of the fundamental properties in nature, which leads to nonlocal correlations between distance objects and has no intuitive explanation in classical physics. In the wake of Bell's theorem \cite{Bell1964}, which shows incompatibility between the local hidden variable (LHV) model and the predictions stipulated by quantum mechanics, there has been a stream of experiments falsifying LHV models by violating the Bell inequality. However, it is only recently we have witness loophole-free confirmation of Bell nonlocality \cite{Hensen2015}, showing the need of further development of quantum technologies to realize quantum entangled-based applications. Entanglement across multiple nodes \cite{Kimble2008,Acin2007,Sangouard2011} promise to enable a wide range of applications like secure communication, distributed quantum computing, cryptography, quantum key distribution, random number generation and quantum sensing \cite{Jiang2007, Broadbent2009,Ekert2014,Nickerson2014,Komar2014}. 
Connection of multiple entangled systems into a network structure is a backbone of future quantum internet \cite{Wehner2018} which, by using entanglement-swapping mechanism \cite{Zukowski1993}, can in principle enable long-distance quantum key generation for secure communication \cite{Ekert1991,Barrett2005}. To realize the potential of quantum internet, it is important to establish reliable certification methods of non-local correlations between nodes in the network. Historically, entanglement swapping protocols based on Event-ready Bell test \cite{Zukowski1993} led to the first bipartite quantum network where conditional successful outcome from an intermediate node together with correlation outcome between terminal nodes is used to disprove an LHV model (see Fig. 1a). Naturally, a general quantum network consists of truly independent sources, hence it is important to investigate a different scenario based on several independent hidden variables \cite{Tavakoli2021}. Recent experiments have considered this scenario to extend LHV model into bilocal hidden variable model (BLHV) proposed by \cite{Branciard2010}, further investigated by \cite{Branciard2012, Tavakoli2014, Gisin2017, Andreoli2017, Tavakoli2020} and experimentally verified by \cite{Carvacho2017,Saunders2017, Sun2019, Poderini2020} (see Fig.1b). However, bilocal inequality violation which disprove BLHV model is not device independent certification of nonlocal sources between terminal nodes \cite{Pozas2021}. It is therefore of principal importance to find a general model which describes a fully non-local network enable to witness nonlocal resources between the end nodes. A solution has been recently proposed by A. Pozas-Kerstjens et al. \cite{Pozas2021} which addresses the shortcoming of BLHV model and lead into a quantum protocol that can be implemented in photonic system for bilocal case scenario. The aim of our work is to experimentally investigate this novel Bell-type inequality tailored for certifying full network non-locality (FNN). Our experiment is implemented by using two pairs of polarized entangled photons in a network configuration with three nodes, referred  shortly as a bilocal network scenario.  Here, two distant parties Alice and Charlie establish a quantum communication channel by independently generating entangled photons. An intermediate party, Bob, performs Bell-state measurement (BSM) using one photon from each entangled pair. The correlations which are generated between all parties lead to violation of the Bell-type inequality, allowing us to disprove any presence of LHV sources in the network. 
\\
\\
In comparison to the standard Bell scenario, new forms of nonlocality arising from different quantum network structures \cite{Branciard2010,Fritz2012,Renou2019,Renou2021} are interesting to investigate due to their importance for future communication and certification protocols. Generally, for a  bilocal network composed of two sources and two communication parties, each of whom selects a private input $x_k$ and produces an output $a_k$,  the correlations follow a network local model if  they can describe  by each source independently emitting a local hidden variables $\lambda_j$  \cite{Pozas2021}:
\begin{align}\nonumber
p(\bar{a}|\bar{x}) &=\int d\lambda_1\mu_1(\lambda_1) \int d\lambda_2\mu_2(\lambda_2)\\
&\times p(a_1|x_1,\bar{\lambda_1}) p(a_2|x_2,\bar{\lambda}_2),
\end{align}
where $\bar{a}=\left \{ a_1,a_2 \right \}$ and $\bar{x}=\left \{ x_1,x_2, \right \}$ are sets of output and  inputs receptively, $\mu_1,\mu_2$ are probability density functions and $\bar{\lambda}_k$ is a set of local variables associated with the source that connect to node $k$. Independence of $\lambda_j$ emerges naturally as a construction of independent source in a network. In its simplest form, with a single source ($\lambda_1=\lambda_2$), the model reduces to original Bell’s notion of LHV \cite{Bell1964,Svetlichny1987}. Extending to two independent sources, we can form a three node network described by two independent hidden variables $\lambda_1$ and $\lambda_2$, thus form a BLHV model \cite{Branciard2010}:
\begin{align}\nonumber
p(a,b,c|x,y,z) &=\int d\lambda_1 d\lambda_2\mu_1(\lambda_1)\mu_2(\lambda_2)\\
&\times p(a|x,\lambda_1)p(b|y,\lambda_1\lambda_2) p(c|z,\lambda_2).
\end{align}  
The bilocal inequalities obtained from BLHV model (2) are proof of network nonbilocality as it rules out any bilocal model. Although it is conceptually novel way to violate Bell-like inequality, simple counterexample shows that it can not certify entanglement between the end nodes of a bilocal network \cite{Saunders2017}. In that case we need to embrace an even stronger concept in notion of network nonlocality where all sources are to be of nonlocal nature.  Introduced by \cite{Pozas2021} we  define full network nonlocality (FNN) if the network cannot be described by allowing at least one source in the network to be of a local nature, while the rest are characterized to be independent nonlocal (see Fig. 1c). Formalizing this concept to bilocal case scenario, the non-FNN correlations are given by
\begin{equation}
p(a,b,c|x,y,z)=\int d\lambda \mu(\lambda) p(a|x,\lambda)p(b,c|y,z,\lambda).
\end{equation}

\begin{figure*}[t]
	\includegraphics[width=0.96\textwidth]{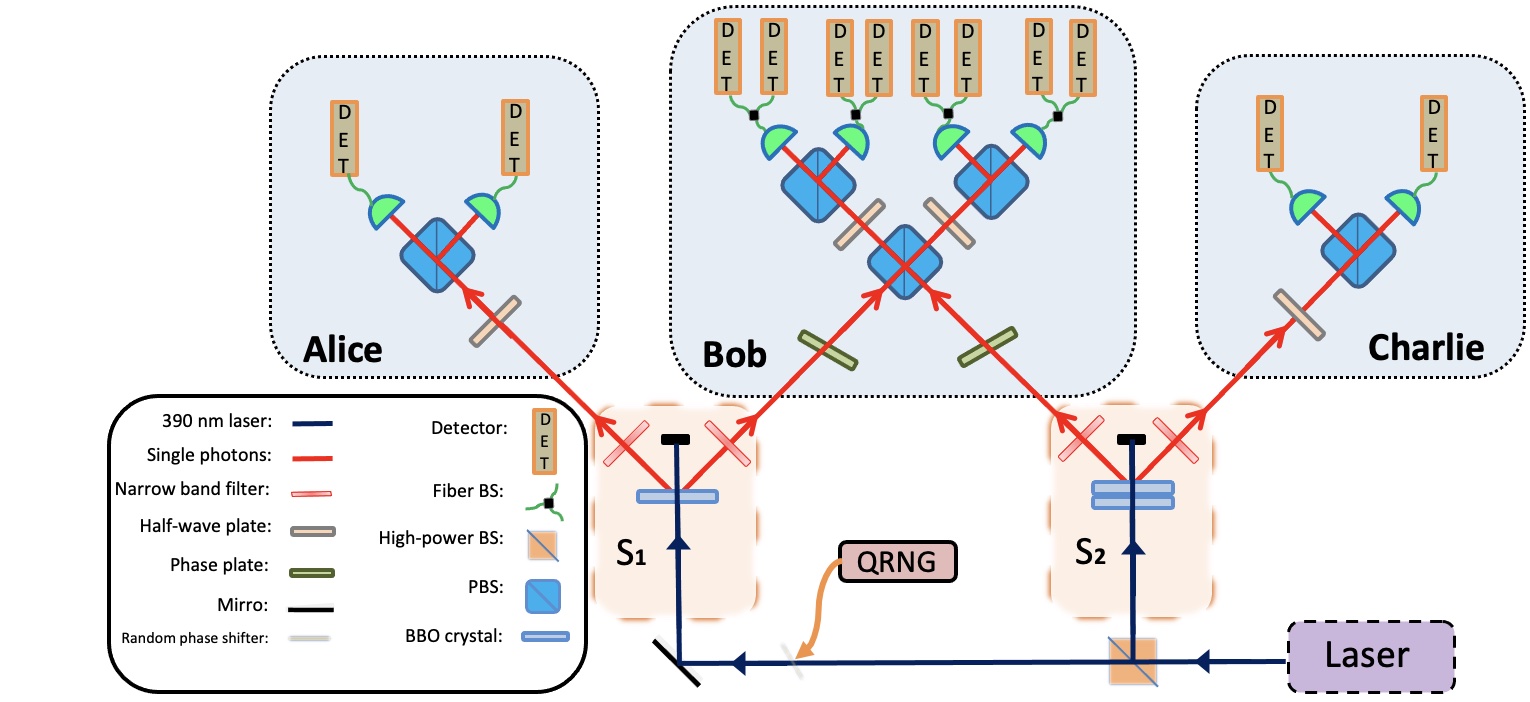}
	\justifying{ \\ \textbf{Experimental setup for testing FNN model in a bilocal network configuration:} The stations of Alice, Bob, and Charlie are highlighted in blue shading and the two entanglement sources S1 and S2, which connect them, are highlighted in orange. Both sources are pumped by ultraviolet light centered at a wavelength of 390 nm with a repetition rate of 80 MHz. Source $S_1$ consist of one BBO crystal and source $S_2$ consist of two BBO crystals in cross-configuration. The spatial, spectral and temporal distinguishability between the down-converted photons is carefully minimized by using narrow band filters, quartz-wedges, and fine alignment translation stages (the last two not included in the picture). To actively remove the coherence between laser pumping into the two sources we implement a quantum random number generator (QRNG) connected to a variable phase shifter (see further in SM). Alice and Charlie implement their measurements using half–wave plates, polarizing beam splitters (PBS) and couplers to single photon detectors (actively quenched Si-avalanche photodiodes, DET) via single mode fibers. Bob implements his partial BSM, which has two input ports, using a single PBS and two half-wave plates. By using fiber 50:50 beam splitter (BS) to split each of the four output ports, Bob is able to do partial photon-number resolving detection and therefore gain information about the two unresolvable Bell states. The observable four-photon coincidence events, which are generated by one simultaneous click for Alice, Charlie and two clicks for Bob, are registered by field programmable gate array and further the  data is processed to calculate conditional probabilities $p(a,b,c|x,y,z)$. Bob’s outcomes; $ b = 0$  corresponds to projection onto the Bell state $|\phi^+\rangle$, $ b = 1$ corresponds to $|\phi^-\rangle$ and  $ b = 2$ corresponds to unresolvable $|\psi^-\rangle$ or  $|\psi^+\rangle$.} \label{Fig2}
\end{figure*}

By considering swapping of the sources, it has been shown that produced conditional distributions, given by $p(a,b,c|x,y,z)$, leads to following FNN witnesses where both need to be violated simultaneously \cite{Pozas2021}: 
\begin{align}\nonumber
\mathcal{R}_{C-NS} &= 2\left \langle A_0B_1C_0 \right \rangle - 2\left \langle A_0B_1C_1 \right \rangle + 2\left \langle A_1B_0C_0 \right \rangle +\left \langle A_1B_0C_1 \right \rangle \\
& -\left \langle B_0 \right \rangle + \left \langle C_1 \right \rangle \left [ \left \langle A_1B_0 \right \rangle + \left \langle B_0C_0 \right \rangle - \left \langle C_0 \right \rangle \right ] \leq 3
\end{align}  
and
\begin{align}\nonumber
\mathcal{R}_{NS-C} &= 2\left \langle A_0B_1C_0 \right \rangle - 2\left \langle A_0B_1C_1 \right \rangle + \left \langle A_1B_0C_0 \right \rangle \\
& \nonumber + 2\left \langle A_1B_0C_1 \right \rangle -\left \langle B_0 \right \rangle 
+ \left \langle A_1 \right \rangle \left  \langle A_1B_0 \right \rangle + \left \langle A_1 \right \rangle \left \langle B_0C_1 \right \rangle \\
& + \left \langle A_1 \right \rangle \left \langle C_0 \right \rangle - \left \langle A_1 \right \rangle \left \langle C_1 \right \rangle - \left \langle A_1 \right \rangle \left \langle A_1 \right \rangle \leq 3, 
\end{align}
where the  expectation values are computed in accordance to \cite{Branciard2012}
\begin{align}\nonumber
\left \langle A_xB_0C_z \right \rangle = \sum_{a,c}-(1)^{a+c} [p(a,0,c|x,z) \\
\nonumber +p(a,1,c|x,z) - p(a,2,c|x,z) ]
\end{align}
and 
\begin{equation} \nonumber
\left \langle A_xB_1C_z \right \rangle = \sum_{a,c}-(1)^{a+c} \left [ p(a,0,c|x,z) - p(a,1,c|x,z) \right ].
\end{equation}

In our experiment the two entanglement sources emit singlet state ($\psi^{-}$). Bob performs a partial BSM, with three different outcomes that are parameterized by ternary bit $\{0,1,2\}\in b$, corresponding  to state  $\phi^+$,  $\phi^-$, and the third corresponding to undistinguishing between the two remaining Bell states ($\psi^+$ and $\psi^-$).  Alice and Charlie have to choose between two possible dichotomic measurements, $\{x,z\}\in \{0,1\}$, corresponding to observables $A_0=\sigma_x$, $A_1=\sigma_z$, $C_0=\tfrac{\sigma_z+\sigma_x} {\sqrt{2}}$ and $C_1=\tfrac{\sigma_z-\sigma_x} {\sqrt{2}}$ and are producing binary outputs $\{a,c\}\in \{0,1\}$.  The  resulting  distribution  leads  to  the  violation of  $\mathcal{R}_{C-NS} = \mathcal{R}_{NS-C} =  5/\sqrt{2} \approx 3.5355$

The schematics of our setup is presented in Fig.2. A femtosecond pulsed laser with repetition rate of 80 MHz producing ultraviolet light (390 nm) which is split and focused into two different setups acting as our two independent sources of entanglement. The first source consists of a crossed configuration of two  2 mm thick  beta barium borate (BBO) nonlinear crystals producing polarization-entangled photon pairs through the process of second order degeneration type-1 spontaneous parametric down conversion (SPDC). The second source consists of one 2 mm thick type-2 BBO crystal generating non-collinear polarization-entangled photons at 780 nm. To further insure independency between the sources we introduce decoherence by  a movable glass plate in one of the pumping pulsed laser paths. The glass plate is steered by a moving motor which receive rotation coordinates from a quantum random number generator, described further in supplementary material. One photon from each pair is sent to a BSM analyzer, where they are interfered. To get a high-quality interference it is necessary for the photons from both paths to be indistinguishable in their temporal, spatial, and spectral modes. To increase spectral overlap we use 3 nm narrowband interference filters (IF). We verify indistinguishability by checking four-photon coincidence in a Hong-Ou-Mandel (HOM) experiment (see supplementary material). The measured HOM dip gave us a visibility value of $v=(89\pm1.3)\%$. We have measured for  each individual source the  visibility in diagonal polarization bases which resulted in visibility of $(99.1\pm0.1)\%$ and $(98.0\pm0.1)\%$ for the sources  1 and 2  respectively. It is well know that  with linear optics one cannot  distinguish between all four Bell states, leading us to distinguish only two of them, $ \phi^{\pm}$. However, by using beam splitters at the outputs of BSM analyzer we can detect, with $50 \%$ probability the presence of other two Bell states, $ \psi^{\pm}$, although without possibility to distinguish them apart \cite{Kwait1998}. The photons were detected with twelve avalanche photodiodes (APD) with an average efficiency of approximately 55-60 \% and the output signals from these detectors sent to a field programmable gate array (FPGA) based  coincidence counting  unit. The total experiment lasted for approximately 15 hours at low pumping power to avoid creation of multiple photons pairs. The produced correlations from the experiment rendered FNN witnesses of $\mathcal{R}_{C-NS}=3.17 \pm 0.05$ and $\mathcal{R}_{NS-C}=3.17 \pm 0.05$, more than three standard deviations above FNN inequality (4) and (5) (see supplementary material for further details). 
\\
\\
In summary, we have for the first time demonstrated an experiment to show FNN for bilocal scenario. The resulted violations of $\mathcal{R}_{NS-C}=3.17 > 3$ and  $\mathcal{R}_{C-NS}=3.17>3$ are well above the classical bound. However, obtained values  are lower than predictions $\mathcal{R}_{NS-C}=\mathcal{R}_{C-NS}\approx 3.5355$ \cite{Pozas2021} can be explained by an imperfect HOM-dip visibility. Similarly to any Bell-type inequalities, our experiment of FNN is subject to detection and locality loopholes. From a more practical perspective, our findings open up for considering complex quantum networks and its connection to FNN. This involves configurations in a hybrid variant where more than two nodes are connected by nonlocal source together with one or several distance nonlocal sources. This will arguably be of great interest from the quantum information point of view as it also provides a certification method to guarantee entanglement between two distance parties just sharing a BSM device.

\section*{Methods}

\textbf{Experimental details.} Correlated photons pair were generated by two separated  and independent SPDC sources. The first source composed of two 2 mm thick BBO nonlinear crystals in a configuration where one crystal was rotated 90 degrees relative to the other. The second source composed of a single 2 mm thick BBO crystal producing non-collinear entangled photons. The pumping fields with wavelength of $\lambda=780$ nm were adapted in power to generate approximately equal amount of correlated photons from each source. After spatial and temporal walkoff compensation and narrowband filtering (10nm at Alice and 3 nm at Bob's and Charlie's) the photons were collected by three different stations using single mode fibers coupled to avalanche photodetectors. At Alice's station, the observable $A_0=\sigma_x$ corresponds to half-wave plate rotated by $\theta^A_0=22.5^\circ$ and $A_1=\sigma_z$ corresponds to half-wave plate rotated by $\theta^A_1=0^\circ$. At Charlie's measuring station $C_0=\frac{\sigma_z+\sigma_x}{\sqrt{2}}$ corresponds to half-wave plate rotated by $\theta^C_0=11.25^\circ$ and $C_1=\frac{\sigma_z-\sigma_x}{\sqrt{2}}$ corresponds to half-wave plate rotated by $\theta^C_0=-11.25^\circ$. Bob's station has only one input, performing a partial BSM which compose of a polarisation beam splitter together with two half-wave plates at $22.5^\circ$. Each measuring configuration takes about 3.5 hours to perform and we use fair-sampling assumption.

\begin{acknowledgments}
We like to thank Dr. Armin Tavakoli for useful discussions. This work was supported by the Knut and Alice Wallenberg Foundation through the Wallenberg Center for Quantum Technology (WACQT) and the Swedish research council.
\end{acknowledgments}	
	
\bibliography{references_qNetworks}

\end{document}